\documentclass{JHEP3}
\title{K-string tensions at finite temperature\\ and integrable models}
\author{Michele Caselle, Pietro Giudice, Ferdinando Gliozzi, Paolo Grinza, Stefano Lottini\\
Dipartimento di Fisica Teorica, Universit\`a di Torino and \\
INFN, Sezione di Torino\\
via P.Giuria 1, I-10125 Torino, Italy\\
\email{caselle,giudice,gliozzi,grinza,lottini@to.infn.it}}
\abstract{It has recently been pointed out that simple scaling properties of Polyakov correlation functions of gauge 
systems in the confining phase suggest that 
the ratios of k-string tensions in the low temperature region is constant 
up to terms of order $T^3$. Here we argue that, at least in a 
three-dimensional $\mathbb{Z}_4$ gauge model, the above ratios 
are constant in the whole confining phase. This result is obtained by 
combining numerical experiments with known exact results on the mass spectrum 
of an integrable two-dimensional spin model describing the infrared behaviour of 
the gauge system near the deconfining transition.}
\usepackage[dvips]{graphicx}
\usepackage{amsfonts,amsmath}
\usepackage{wrapfig}
\usepackage{mathrsfs}
\usepackage{amssymb}
\usepackage{psfrag}
\usepackage{epsfig}
\usepackage{psfrag}
\newlength{\intwidth}

\newcommand{\EQ}{\begin{equation}}
\newcommand{\EN}{\end{equation}}
\newcommand{\bea}{\begin{eqnarray}}
\newcommand{\eea}{\end{eqnarray}}

\newcommand{\Z}{\mathbb{Z}}

\newcommand{\R}{\mathscr{R}}
\newcommand{\U}{\mathscr{U}}
\newcommand{\bra}{\langle}
\newcommand{\ket}{\rangle}
\newcommand{\avg}[1]{\langle \hspace{0.2em} #1 \hspace{0.2em} \rangle}

\newcommand{\sun}{\mathop{\rm SU}(N)}

\newcommand{\eq}{\begin{equation}}
\newcommand{\en}{\end{equation}}
\newcommand{\ea}{\end{eqnarray}}

\newcommand{\link}[1]{\bra #1\ket}

\begin{document}

\section{Introduction}
% pstoedit -f fig eps.eps > eps.fig
In most confining  gauge theories, besides the fundamental string (of tension $\sigma$) which is formed between 
a pair of static sources in the fundamental representation $f$, there is the freedom of taking the sources in
 any representation $\R$. 

If, for instance, the gauge group is $\sun$ there are infinitely many 
irreducible representations at our disposal. However, as the sources are
pulled apart, no matter what
representation is chosen, the asymptotically stable string tension 
$\sigma_{\R}$  depends only 
on the $N-$ality $k$ of $\R$, i.e. on the number (modulo $N$) of copies of 
the fundamental representation needed to build $\R$ by tensor product, 
because
all representations with the same $k$ can be transformed into each other
by the emission of a proper number of soft gluons. As a consequence the
heavier strings decay into the string of smallest string tension $\sigma_k$. 
The corresponding string is referred to as a k-string. This kind of confining object can be defined whenever 
the gauge group admits more than one non trivial 
irreducible representation.

Much work has been done in the study of k-string tensions in the continuum \cite{ds}--\cite{Armoni:2006ri} as 
well as
on the lattice \cite{lt1}--\cite{ddpv}. 

In a previous work \cite{confanomaly}, some of us have argued from simple 
scaling properties of suitable Polyakov loop correlators that these string 
tensions have the following 
low temperature asymptotic expansion 
\EQ
\label{tension}
	\sigma_k(T) = \sigma_k - c\frac{\pi}{6}T^2 + \mathcal{O}(T^3) \; ; \; 
	c = (d-2)\frac{\sigma_k}{\sigma	} \; ,
\EN
where $c$ is the central charge of the underlying 2D conformal field theory 
describing the IR behaviour of the k-string. As a consequence, their ratios are expected to be constant up to 
 $T^3$ terms:
\EQ
\frac{\sigma_k(T)}{\sigma(T)}=
\frac{\sigma_k}{\sigma} + \mathcal{O}(T^3) \,\,.
\label{ratios}
\EN   

The low temperature data presented in support of this expectation were 
taken from Monte Carlo simulations on a particular system, namely a 
(2+1)-dimensional $\Z_4$ gauge model, which is the simplest 
exhibiting more than just the fundamental string. 

The main conjecture we want to verify in this work is that 
$\sigma_k(T)/\sigma(T)$, at least in that $\Z_4$ gauge system, is in fact 
independent of the temperature in the \emph{whole} of the confining regime. 
To check this idea, a handy fact comes useful, namely that,
as the system approaches the deconfinement transition, and the 
string picture begins fading, another approach is made available by the 
Svetitsky-Yaffe (SY) conjecture \cite{sy}, which allows to reformulate the 
system in a totally different perspective, based on a two-dimensional 
integrable theory 
in which, however, the near-$T_c$ counterpart of the low-temperature 
result cited above can be nicely found.

 It turns out that the deconfinement transition of the 3D 
$\Z_4$ gauge model is second order and, according to the SY conjecture,
  belongs to the same universality class of the 2D symmetric 
Ashkin-Teller (AT) model. As a matter of fact, such a model possesses a 
whole line of critical points along which the critical exponents vary 
continuously. The SY conjecture tells us that if a (2+1)-dimensional gauge 
model with center $\Z_4$   displays a second-order transition, then its universality class is associated to a 
suitable point of the critical line of the 
2D AT model. For instance, it has been argued \cite{su4deforcrand} that the critical (2+1)D $SU(4)$ gauge theory 
belongs to the universality class of a 
special point of the AT model, known as the  four-state Potts 
model. More generally, the class of models with gauge group $\Z_4$ depends 
on two coupling constants $\alpha$ and $\beta$, and the universality 
class of the deconfining point $P$ varies with the ratio $\alpha/\beta$.

The two-dimensional AT model can
be seen in the continuum limit as a bosonic conformal field theory plus a 
massive perturbation  (i.~e.~a Sine-Gordon theory)
driving the system away from the critical line. 
Thus, a map between (a neighbourhood of) the AT critical line 
and the Sine-Gordon phase space is provided.

This theory is integrable, and the masses of its lightest physical states 
(first soliton and first breather mode, of masses $M$ and $M_1$) 
correspond to the tensions 
$\sigma(T)$ and $\sigma_2(T)$ near $T_c$, whose ratio, in this context, 
can be analytically 
evaluated and turns out to be
\eq
\lim_{T\to T_c}\frac{\sigma_2(T)}{\sigma(T)}=
\frac{M_1}{M}=2\sin\frac\pi2(2\nu-1)\;\;,
\label{main}
\en
where $\nu$ is the thermal exponent in two dimensions.

As a consequence, on the gauge side, we have two different ways to verify the conjecture. One is to  directly 
estimate the ratio $M_1/M$ by measuring the Polyakov-Polyakov correlators in the two non-trivial representations
 of $\Z_4$.
The other is to evaluate the thermal exponent of the gauge system at the
deconfining temperature. Either method gives a value of ${M_1}/{M}$ which 
nicely agrees with the ratio $\sigma_2/\sigma$ evaluated at $T=0$ .

\subsection{The (2+1)D $\Z_4$ gauge model and its dual reformulation}
\label{dualref}

The most general form of $\Z_4$ lattice gauge model admits two independent coupling constants, with partition 
function
\EQ
	\Z(\beta_f,\beta_{ff}) = \prod_l \sum_{\xi_l = \pm1, \pm i} e^{\sum_p (\beta_f \U_p + \beta_{ff} \U_p^2/2
	 + \mathrm{c.c.})} \; ,
\EN
in which the gauge field $U_l$ on the links of a cubic lattice is valued among the fourth roots of the unity 
and the sum in the exponent is taken over the elementary plaquettes of the lattice.
Such a theory can be reformulated as two coupled $\Z_2$ gauge systems:
\EQ
	\Z(\beta_f,\beta_{ff}) = \prod_l \sum_{\{ U_l=\pm 1, V_l = \pm 1 \}} e^{\sum_p [ \beta_f(U_p+V_p) + 
	\beta_{ff} U_p V_p ] } \;\;\;, 
(U_p=\prod_{l\in p}U_l~~;~~V_p=\prod_{l\in p}V_l\;).
\EN

 From a computational point of view it is useful to exploit a duality relation, switching to a spin model, that 
in this case is a 3D AT model, expressed as a 
double Ising spin field plus a coupling term between the Ising variables $\{\sigma\}$ and $\{\tau\}$:
\EQ
	\label{eq:3dataction}
	S_{AT}(\alpha,\beta) = - \sum_{\link{xy}} [\beta(\sigma_x\sigma_y+\tau_x\tau_y)+\alpha(\sigma_x
	\sigma_y\tau_x\tau_y)] \; ;
\EN
the duality is implemented by
\bea
		\alpha & = & \frac{1}{4} \ln \Big[ \frac{(\coth\beta_f+\tanh\beta_f\tanh\beta_{ff})
			(\coth\beta_f+\tanh\beta_f\coth\beta_{ff})}
			{2+\tanh\beta_{ff}+\coth\beta_{ff}} \Big] \; , \\
		\beta & = & \frac{1}{4} \ln \Big[ \frac{1+\tanh^2\beta_f\tanh\beta_{ff}}
			{\tanh^2\beta_f+\tanh\beta_{ff}} \Big] \,\,  .
\ea

The choice of working in the dual spin version of the gauge system is strongly motivated by the availability 
of highly efficient nonlocal Monte Carlo algorithms, in which, moreover, any kind of gauge-invariant observable 
can be directly embedded in the update procedure.

The $\Z_4$ gauge model admits, in addition to the fundamental string, a
 k-string with $k=2$, corresponding to taking the sources in the 
double-fundamental representation $f\otimes f$.

The phase diagram of this 3D model at $T=0$ has been studied long ago
\cite{dbgk,az}. The deconfinement transition, which is weakly first order in 
the region we are interested in, becomes second order at finite $T$, therefore, according to SY conjecture, is 
described by the order-disorder transition of
a 2D AT model.      
As anticipated in the Introduction, such a transition forms a whole 1-dimensional manifold of  critical points 
(see Figure \ref{fig:atphasespace}). 
Along this line, the critical indices (and other universal quantities as well)
 vary continuously, the endpoints representing a decoupled double 
Ising system and the 4-state Potts model. 
The choice of the point in the phase space in which to work is therefore 
a crucial issue.

\FIGURE{
\centering
\includegraphics[angle=0, width=10cm]{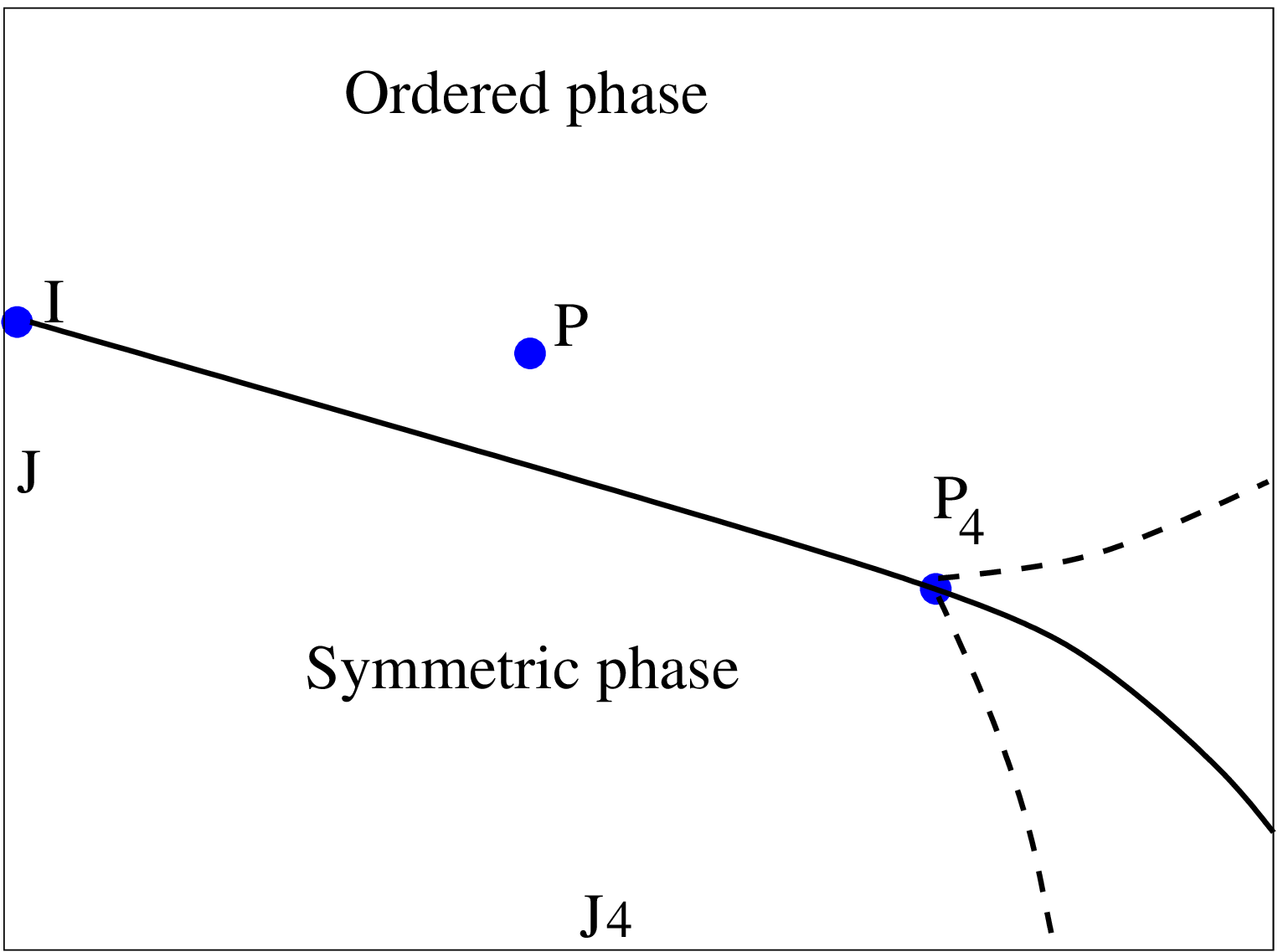}
\caption{Sketch of the phase diagram of the Ashkin-Teller model in 
two dimensions (see Eq.~\ref{eq:2dataction} for the meaning of the 
constants $J$, $J_4$). The point labelled $I$ corresponds to a pair of 
decoupled critical Ising systems 
($J_4 = 0$, $J = J_c^{\mathrm{Ising}}$), while 
the $P_4$ point ($J_4=J=J_c^{\mathrm{Potts}}$) 
represents the four-state Potts critical point. The solid (self-dual) 
line between them is the one-dimensional critical manifold along which 
the critical indices vary with continuity. Beyond $P_4$, the 
self-dual line is no longer critical, and another phase 
appears (bounded by the two dashed lines departing from $P_4$). 
The point $P$ indicates the image of the confined three-dimensional 
gauge system near $T_c$ according to the Svetitsky-Yaffe conjecture.}
\label{fig:atphasespace}
}

 From the data in \cite{confanomaly}, obtained by means of 
finite-temperature measurements of  Polyakov-Polyakov correlation functions, and particularly 
from those referring to the point $P$ identified by $(\alpha,\beta)=(0.050, 0.207)$, the string tensions 
$\sigma$ and $\sigma_2$ can be evaluated in the $T \to 0$ 
limit as temperature-independent quantities:
\bea
	\sigma\,a^2 & = &   0.02085(10) \nonumber \; ,\\
	\sigma_2\,a^2 & = & 0.03356(22)  \; ,
\ea
where $a$ is the lattice spacing.
Their ratio, which has been argued to equate the central charge of the CFT related to the 2-string, is then 
given by
\EQ
	\frac{\sigma_2}{\sigma} = 1.610(13) \; .
\label{ratiosigma}
\EN

% FINE STEO
% INIZIO PAOLO

\section{The Svetitsky-Yaffe conjecture and the Sine-Gordon model}
% \subsection{The scaling field theory}
The mapping induced by the Svetitsky-Yaffe conjecture leads to a substantial simplification in the study 
of the critical properties of the deconfining transition, allowing to study it as a standard symmetry-breaking 
transition which takes place in a spin model. In the present case we 
deal with the symmetric Ashkin-Teller model in two-dimensions.

The action for this model has the same form of Eq.~(\ref{eq:3dataction}), 
but to clarify the fact that this is the 2-dimensional 
AT system reached via SY conjecture (as opposed to the three-dimensional 
one dual to the original gauge theory), we will relabel the spin fields and coupling constants with other 
names within this context:
\EQ
	\label{eq:2dataction}
	{\cal{S}}_{AT} = - \sum_{\link{xy}} [J(\sigma^1_x\sigma^1_y+\sigma^2_x\sigma^2_y)+J_4(\sigma^1_x
	\sigma^1_y\sigma^2_x\sigma^2_y)] \; .
\EN

Such a model has been extensively studied in the past, and a number of analytic and numerical results have
 been discovered 
\cite{dbgk,baxter}. Its phase diagram is exactly known (Fig.~\ref{fig:atphasespace}).  

The critical line is characterised by the fact that it is self-dual 
and separates a disordered phase from an ordered one. 
Since the symmetry which is spontaneously broken by crossing it 
is the global $\mathbb{Z}_4$ symmetry of the model, this 
is also the critical manifold corresponding to  the deconfining phase 
transition of the gauge model. 
Hence in the following we will concentrate on (a part of) this critical line.  

Important analytic results related to the critical line have been worked out by a direct solution of the lattice
 model \cite{baxter}. 
Another interesting approach was carried on in the paper \cite{kadanoff}, by considering the $c=1$ conformal 
field 
theory describing the critical line, i.e.~the Gaussian model. In such a context it was possible to establish an 
exact correspondence 
between lattice/continuum operators and to compute exactly their 
conformal dimensions.

The advantage of working in the field theoretical setting is that 
one has the possibility to study the off-critical behaviour near 
the critical line in a very natural way. This is of great convenience for us, since we will be ultimately 
interested in the mass 
spectrum of the theory in the high temperature phase.
    
Let us briefly sketch how to obtain a field theoretic description of the scaling region near 
the critical self-dual line of the model. Since the correlation length remains larger than the 
lattice spacing, the model can be described by the following QFT:
\bea
{\mathcal A_{\textrm{\tiny AT}}} = {\mathcal A_{\textrm{\tiny IM}}^{(1)}}+
{\mathcal A_{\textrm{\tiny IM}}^{(2)}}+ \tau \, \int d^2 x \, ( \epsilon_1(x) +\epsilon_2(x) ) \; +
\rho \,   \int d^2 x \,  \epsilon_1(x) \, \epsilon_2(x)  \; ,
\label{thermal}
\eea
where the meaning of such an expression is quite transparent when 
compared to the Hamiltonian of the lattice model. 
The first two terms stand for the conformal field theories with $c=1/2$ describing the critical behaviour of the 
two Ising models, and the latter two terms are respectively the relevant thermal perturbation ($\epsilon_i(x)$
 are 
the energy operators in the two copies of the Ising model), and the marginal one which moves the system along 
the 
critical line. It is also evident by comparison that the couplings $\tau$ and $\rho$ are substantially a 
reformulation 
of the former $J$ and $J_4$ respectively.
  
In other words, when $\tau=0$, $\rho \neq 0$, the critical line is described by a compactified free massless 
boson (Gaussian model). A full analysis of the critical line of the Ashkin-Teller model by means of the Gaussian
 model was established in \cite{kadanoff}. 

The general Hamiltonian with $\tau \neq 0$ describes the model outside the critical line. 
It is useful to note that it can be seen as a perturbation of the Gaussian model, and in such a bosonic language 
the thermal perturbation can be written as $\cos \beta \varphi$, where $\beta$ is a marginal parameter 
equivalent to $\rho$. Hence we are left with
\bea
{\mathcal A_{\textrm{\tiny AT}}} = \int d^2 x \, \left( \frac12 \partial_\mu \varphi \partial^\mu \varphi \, - \, 
\tau \,  \cos \beta \varphi
\right) \; ,
\eea
which is the action of the Sine-Gordon model. In this notation, we have the high-temperature phase for $\tau>0$,
 and the low-temperature one for $\tau<0$.
It was also shown that the previous model describes the Ashkin-Teller model in the range $2 \pi \le \beta^2 \le 
6 \pi$. 
Actually we are interested in the narrower range $2\pi\le \beta^2\le 4\pi$ 
from the critical 4-state Potts model to two decoupled critical Ising models.
 Furthermore, since the confined phase of the gauge theory is mapped in the high-T phase of the Ashkin-Teller model, we 
 will only consider the case $\tau>0$. 

Such a QFT is of particular interest because it is integrable, and this is 
the main reason for rewriting the action of 
the model near the critical point in a bosonic form\footnote{It is always possible to \emph{fermionise} the 
action of the Sine-Gordon model in order to obtain an integrable fermionic theory with the same scattering 
matrix, namely the massive Thirring model.}. 

Integrability means that an infinite number of integrals of motion exists. The main consequence in (1+1)-dimensions is the fact that the scattering theory is
very constrained, because the $S$-matrix is factorised in products of two-body interactions, and inelastic 
processes are forbidden. 

These facts allow to write down the so-called Yang-Baxter 
equations for the 2-particle $S$-matrix. Then, such an $S$-matrix can be computed exactly by imposing the 
previous equations and the usual requirements of unitarity and crossing (for a review about Integrable QFTs 
see \cite{revgius}). 

An obvious consequence is that also the spectrum of the masses of the bound states of the theory is known 
exactly, since they are represented by the simple poles of the $S$-matrix in the physical strip.

It is worth to recall that an important consequence of the Svetitsky-Yaffe conjecture is that the ratio of 
a given string tension over the temperature of the gauge theory is 
mapped, near $T_c$,
 onto a corresponding mass of the spin model spectrum.
Then, the ratio of string tensions $\sigma_2/\sigma$ is mapped onto a suitable ratio of masses of the 
Sine-Gordon model which is known exactly as a function of universal quantities.

In the following we will make some quantitative considerations about the qualitative picture given above in 
order to make some predictions which will be useful in the context of the gauge theory. Since a detailed analysis 
of the properties of the scattering theory of the Sine-Gordon model in the context of the Ashkin-Teller model 
has been done in \cite{aldopaolo}, we will refer to those papers for the details.

\subsection{Operators correspondence, mass spectrum and correlation functions}
\label{sub-operator}

The last ingredient we need before exploiting the map to the Sine-Gordon model at its best is the 
correspondence between the Polyakov loops in higher representation and the operators of the Ashkin-Teller model.

We already know from the Svetitsky-Yaffe original work that the Polyakov loop 
in the fundamental representation corresponds to the spin operator. 
Then, following the same reasoning used in \cite{kdk}, it is possible to deduce that the Polyakov loop in the 
double fundamental representation is related to the so-called \emph{polarisation} operator ${\mathcal P} = 
\sigma^1  \sigma^2$,
where $\sigma^1$ and $\sigma^2$ are the spin variables defined in 
(\ref{eq:2dataction}). Its bosonic form and the corresponding  anomalous 
dimensions are given by
\bea
{\mathcal P} = \sin \frac{\beta}{2} 
\varphi, \ \ \ \ \ X_{\mathcal P} = \frac{\beta^2}{8 \pi} \; ;
\eea
we also notice that $\langle {\mathcal P} \rangle =0 $ in the 
high-T phase of the model.

\underline{Sine-Gordon mass spectrum \cite{refsuSGsmatrix}:} The exact 
knowledge of the S-matrix allows to access to the exact mass spectrum of the theory. Without entering into the 
details, the spectrum of the SG model is given by a soliton/anti-soliton doublet of fundamental particles 
of mass $M$, and a number of soliton/anti-soliton bound states, called breathers $B_n$, whose number is a 
function of $\beta^2$.
By defining the coupling constant $\xi$ in the following way
\bea
\xi = \frac{\pi \, \beta^2}{8 \pi - \beta^2} \; ,
\eea
we have that for $\xi \ge \pi$, i.e. $\beta^2 \ge 4 \pi$, no bound states 
are present and hence the spectrum is given by the soliton/anti-soliton 
doublet only (repulsive regime).

For $\xi < \pi$, i.e. $\beta^2 < 4 \pi$, we are in the attractive regime 
and the breathers $B_n$ appear as simple poles of the S-matrix. 
Their number and masses are given by the following formula
\bea
M_n \ = \ 2 M \, \sin \frac{n}{2} \, \xi\, , \ \ \ \ 1 \le  n < \left[ \frac{\xi}{\pi} \right] \; .
\eea
Since we are interested in the range $2 \pi \le \beta^2 \le 4 \pi$, 
we immediately realise that, outside the two decoupled Ising point 
at $\beta^2 = 4 \pi$, we always have at least one breather of 
mass $M_1$ (for $2 \pi \le \beta^2 \le 8/3 \pi$ we also have the 
breather $M_2$ which is however irrelevant for our analysis).   

The next step is to associate particle states to operators in the 
high temperature phase. It has been done in \cite{aldopaolo} by 
taking into account their properties of symmetry and locality. 
 The result is that the spin operator only couples to particle states 
with topological charge equal to one, i.e. involving an odd number of solitons (or antisolitons), and couples to
 even-labelled breathers only. On the contrary the polarisation operator couples to neutral particle states only, 
 i.e.~states with the same number of solitons/antisolitons and odd-labelled breathers. 

As a consequence the spin operator is naturally associated to the mass of the soliton, and the polarisation 
operator is associated to the mass $M_1$ of the breather $B_1$. This means that the string tension of the 
Polyakov loop in the fundamental representation corresponds to the mass of the soliton, and the the string 
tension of the double fundamental corresponds to the first breather.
Hence, following the Svetitsky-Yaffe conjecture, the ratio of string tensions in the confining phase near the 
transition is given by
\bea
\frac{M_1}{M} \ = \ 2 \, \sin \frac{\xi }{2} \; .
\label{eq:ratiomm}
\eea
This result, being a dimensionless ratio, is expected  to be universal in 
the limit $\tau \to 0$. This fact can be explicitly seen by expressing the 
coupling 
$\xi$ in terms of some critical exponent. Not surprisingly, it is indeed 
possible because the theory is solved also at the critical point in terms 
of the Gaussian model. By comparing the power-like behaviour of the 
energy-energy correlator of the Gaussian model:
\bea
\langle \epsilon(x) \epsilon(0) \rangle_{\textrm{\tiny Gaussian}} \ \propto \ \frac{1}{|x|^\frac{\beta^2}{2 
\pi}} \; ,
\eea
with that expected from scaling theory,
\bea
	\label{eq:fss_nu}
	\langle \epsilon(x) \epsilon(0) \rangle_{\textrm{\tiny Gaussian}} \ \propto \ \frac{1}{|x|^{2(d-\frac{1}
	{\nu})}} \; ,
\eea
it is possible to work out the following relation between $\xi$ and the thermal critical exponent $\nu$ (we have
 $d=2$):
\bea
	\xi \    = \pi \, (2\nu-1) \; .
\eea
It yields
\bea
	\label{eq:ratio_nu}
	\frac{M_1}{M} \ = \  2 \, \sin \frac{\pi}{2}  \, (2\nu - 1) \;.
\eea
Such a result is very important because it gives an exact prediction for the ratio $\sigma_2(T) / \sigma(T)$ 
near the deconfining point at $T_c$ as a 
function of the critical exponent $\nu$.

\underline{Large distance behaviour of correlators:} The previous analysis of the mass spectrum allows to compute the 
leading behaviour of the correlators $\langle \sigma \sigma \rangle $ and $\langle  {\mathcal P}   {\mathcal P} 
 \rangle$ at large distance by means of their spectral expansion over form factors (the interested reader can
  refer to \cite{smiyuzam} for the details).
On general grounds, the leading behaviour at large distance is expected 
to obey an exponential decay involving the mass of the lightest state allowed by symmetry and locality.

 The analysis of the previous section immediately allows to write down the leading term for $\langle \sigma 
 \sigma \rangle $ and $\langle  {\mathcal P}   {\mathcal P} \rangle$ correlators in the high-T phase of the 
 theory, up to a proportionality constant
%\bea
%\langle \sigma (x) \sigma (0) \rangle & \sim  & K_0 (M |x|)\; , \ \ \ \ \ |x| \to \infty \nonumber \; ; \\
%\langle  {\mathcal P} (x)   {\mathcal P} (0) \rangle & \sim  & K_0 (M_1 |x|)\; , \ \ \ \ \ |x| \to \infty \; ,
%\label{uu1}
%\eea
				\eq
					\label{uu1}
					\left. \begin{array}{rclc}
							\langle \sigma (x) \sigma (0) \rangle & \sim & K_0 (M |x|)\; , \ \ \ & \ \ |x| \to \infty \; ; \\
							\langle  {\mathcal P} (x)   {\mathcal P} (0) \rangle & \sim  & K_0 (M_1 |x|)\; , \ \ \ &  \ \ |x| \to \infty \; ,
					\end{array}
					\right.
				\en

where $K_0$ denotes the modified Bessel function of order zero, 
and $M$, $M_1$ are the masses of the soliton and the first breather 
respectively. It is interesting to notice that the spectral expansion 
gives the exact asymptotic form of the correlator, and not a generic 
exponential decay. 

An important consequence of this observation is that in the regime in which the Svetitsky-Yaffe correspondence
holds we expect the same large distance behaviour for the effective
string correction (i.e. the term $\frac12 \log{R}$) for both the fundamental and the excited string. 

Let us explain this point in more detail.
The effective string correction in the case of the cylindric geometry of the Polyakov loop correlators has two
very different regimes: for distances $R$ between the Polyakov loops smaller than $L/2$ (where $L$ denotes the
length of the lattice in the compactified time direction) the correction is the usual L\"uscher term
proportional to $1/R$. On the contrary for $R>L/2$, it is given by an universal logarithmic correction:
$\frac12\log{(\frac{2R}{L})}$
 (see for instance eq.~(10) of ref.~\cite{chp02}).

This is the regime (small $L$, i.e.~high $T$) 
in which we may expect the dimensional reduction picture to hold,
and, according to the identification between 2D spin model and 3D gauge theory observables discussed above, this
implies that the prefactor in front of the exponential decay of the two point correlators must be $1/\sqrt{R}$.
This is exactly the prefactor of the $K_0$ function and this coincidence represents a non trivial test
of the reliability of the dimensional reduction program (see the discussion in Sect.~2.2 of~\cite{bc05}).

What is remarkable in the result of Eq.~(\ref{uu1}) is that this same large distance effective string correction
holds unchanged both for the fundamental and for the excited string. This represents a strong constraint for any
consistent effective string model for excited k-strings and is one of the exact predictions on the k-string
behaviour that we can extract from our dimensional reduction analysis.

\vskip0.7cm

Let us summarise the results we discussed in this section in the perspective of applying them to the evaluation 
of the ratio of string tensions near the critical line:
\begin{enumerate}
\item{The exact large distance asymptotic behaviour of the correlators of $\sigma$ and $ {\mathcal P}$ can be 
considered as a reliable tool to extract the lightest mass which governs their exponential decay. 
One can separately compute the Polyakov-Polyakov correlators in the representations $f$ and $f\otimes f$ with a
Monte Carlo simulation and then fit the data in order to extract 
$M$ and $M_1$. Their ratio $M_1/M$ is an estimate of the ratio of the 
string tensions $\sigma_2/\sigma$ near the deconfining transition, and can be directly compared with its estimate 
at zero temperature in order to 
confirm or reject our conjecture. 
}
\item{An independent way to compute the ratio $M_1/M$ is to explicitly 
use the mass formula as a function of the thermal exponent $\nu$ of the gauge theory which, according to the SY 
conjecture, coincides with that of the corresponding 2D model. 
In particular one can study the finite size behaviour of the plaquette 
operator (or the susceptibility) and extract the corresponding value of $\nu$. 
Then by plugging it in the mass formula (\ref{eq:ratio_nu}) one gets 
another independent estimate of  
$M_1/M$, and again it can be compared to the corresponding estimate 
of $\sigma_2/\sigma$ at zero temperature. This is also a direct check 
that the ratio of string tensions follows the proposed analytic formula.
} 

\item{In the large distance regime $L<2R$ both the fundamental and the excited string should be affected by
the same effective string correction: $\frac12\log{R}$}.

\end{enumerate}

\subsection{Baryon vertices and mass spectrum}
\label{baryon}
The general principle invoked in \cite{confanomaly} to derive Eq.~(\ref{tension}) is simply that in a 
d-dimensional gauge theory whatever 
correlation function made with Polyakov loops in the 
\underline{fundamental} representation should be described, 
at sufficiently low temperature and in the IR limit, by a two-dimensional 
conformal field theory with central charge $c=d-2$.
    
A simple consequence of this general principle is that, as long as the 
temperature is far from 
the critical one, the shape of the world-sheet spanned by the baryon 
vertices should be temperature independent; this ensures that the 
baryon static potential has the expected asymptotic form \cite{confanomaly}.
Depending on the location of external sources some fundamental strings 
contributing to the baryon vertex may coalesce, giving raise to the k-string 
formation.  
As noticed in \cite{confanomaly}, the balance of the string tensions 
for a given vertex gives the following expression for the angles 
at the center of the junction of three arbitrary k-strings
\bea
\label{angoli}
\cos \, \vartheta_i = \frac{\sigma^2_j (T) + \sigma^2_k (T) - \sigma^2_i (T)}{2\, \sigma_j (T)\sigma_k (T)}, 
\ \ \ \ \textrm{and cyclic permutations of 
the indices.}
\eea
The rigidity of the geometry of the vertex is then ensured by requiring that such angles are kept fixed when 
the temperature varies. 
As a consequence, all the string tension ratios are constant up to a given 
order in $T$, namely as far as the effective string picture is valid. 
In other words, the previous geometrical construction is likely to 
break down when the system approaches the deconfining temperature, 
as the string begins to fluctuate wildly.

We can summarise the above consideration by saying that in the low 
temperature region the trajectory described in the phase space by the 
gauge system while varying the temperature $T$ is a line of constant physics,
i.e. $\sigma_2(T)/\sigma(T)$ is constant. 

A similar  picture emerges when studying the gauge system near the 
deconfining transition. In the framework of the Svetitsky-Yaffe conjecture 
the second order phase transition of the gauge system is described 
by the critical behaviour of a certain 2D spin model.
Then, the off-critical scaling region of the latter can be described by a 
suitable (1+1)D QFT
\footnote{For a recent application of this approach to 3D $\Z_3$ gauge theory
see \cite{z3potts}.}. Focusing on its scattering properties, it is possible 
to argue that the S-matrix is characterised by its analytic properties 
\cite{eden}.
In particular the two-particle elastic scattering matrix is a multivalued 
function of the Mandelstam variable $s$ ($\theta$ is the rapidity 
which parametrise momentum and energy)
\bea
s=m_1^2 + m_2^2 + 2 m_1 m_2 \cosh (\theta_1 - \theta_2) \; ,
\eea
with a Riemann sheet possessing three branch points (it is otherwise meromorphic), called physical sheet. 
The branch points are at $(m_1+m_2)^2$, $(m_1-m_2)^2$ and $\infty$, and the cuts are located on the real 
line avoiding the interval $[(m_1-m_2)^2$, $(m_1+m_2)^2]$, which is the interval where bound state poles 
can appear (physical strip). Their masses are eventually given by
\bea
m_b^2 = m_1^2 + m_2^2 + 2 m_1 m_2 \cos  u_{12}^b\; , \; \; \; \textrm{triangle of masses} \; ,
\label{triangle}
\eea
where $\theta = i \, u_{12}^b$ is the purely imaginary value of the 
rapidity corresponding to the creation of the particle $m_b$.
 
Let us stress that this is merely a consequence of the kinematics, and it is a 
generic situation for any (1+1)D QFT. It is  nice to see that we are left 
with the very same structure of  angles as Eq.~(\ref{angoli}), which 
is translated 
in the so-called ``triangle of masses'' for the bound state particle $m_b$.
Interestingly, since such masses play the role of string tensions,
we see that they obey the same relation both near $T=0$ and near $T=T_c$, see Fig. \ref{fig:baryon_mass}.
      
Coming back to the case analysed in this paper, we can add a 
further important element to this picture. As explained in the previous 
Sections, the 3D $\Z_4$ lattice gauge theory is mapped 
via Svetitsky-Yaffe to the 2D Ashkin-Teller model near the self-dual 
critical line, which is in turn described by the Sine-Gordon field 
theory. 

One of the main consequences of the integrability of the latter 
is the exact knowledge of the mass spectrum. In the present case the 
 process of coalescence of two fundamental strings into a 2-string 
corresponds to the scattering of a soliton/anti-soliton pair creating 
the bound state $B_1$.
 
\FIGURE{ \centering 
\psfrag{Baryon vertices}{\huge{\textbf{Baryon vertices}}}
\psfrag{Bound states in integrable QFTs}{\huge{\textbf{Bound states in 
integrable QFTs}}}
\psfrag{s_i}[b]{\huge{$\sigma_i(T)$}}
\psfrag{s_j}[b]{\huge{$\sigma_j(T)$}}
\psfrag{s_k}{\huge{$\sigma_k(T)$}}
\psfrag{t_k}{\huge{$\vartheta_k$}}
\psfrag{Balance}{\huge{``Balance of string tensions''}}
\psfrag{eq_1}{ \huge{$\cos \, \vartheta_k = \frac{\sigma^2_i (T) + \sigma^2_j (T) - \sigma^2_k (T)}{2\, \sigma_i (T)\sigma_j (T)}$}}
\psfrag{m_1}{\huge{$m_1$}}
\psfrag{m_2}{\huge{$m_2$}}
\psfrag{m_b}{\huge{$m_b$}}
\psfrag{t_12}{\huge{$\theta=i u_{12}^b $}}
\psfrag{Triangle}{\huge{``Mass triangle''}}
\psfrag{eq_2}{\huge{$m_b^2 = m_1^2 + m_2^2 + 2 m_1 m_2 \cos  u_{12}^b$}}
\includegraphics[angle=0, width=14cm,bb=60 40 800 345,clip]{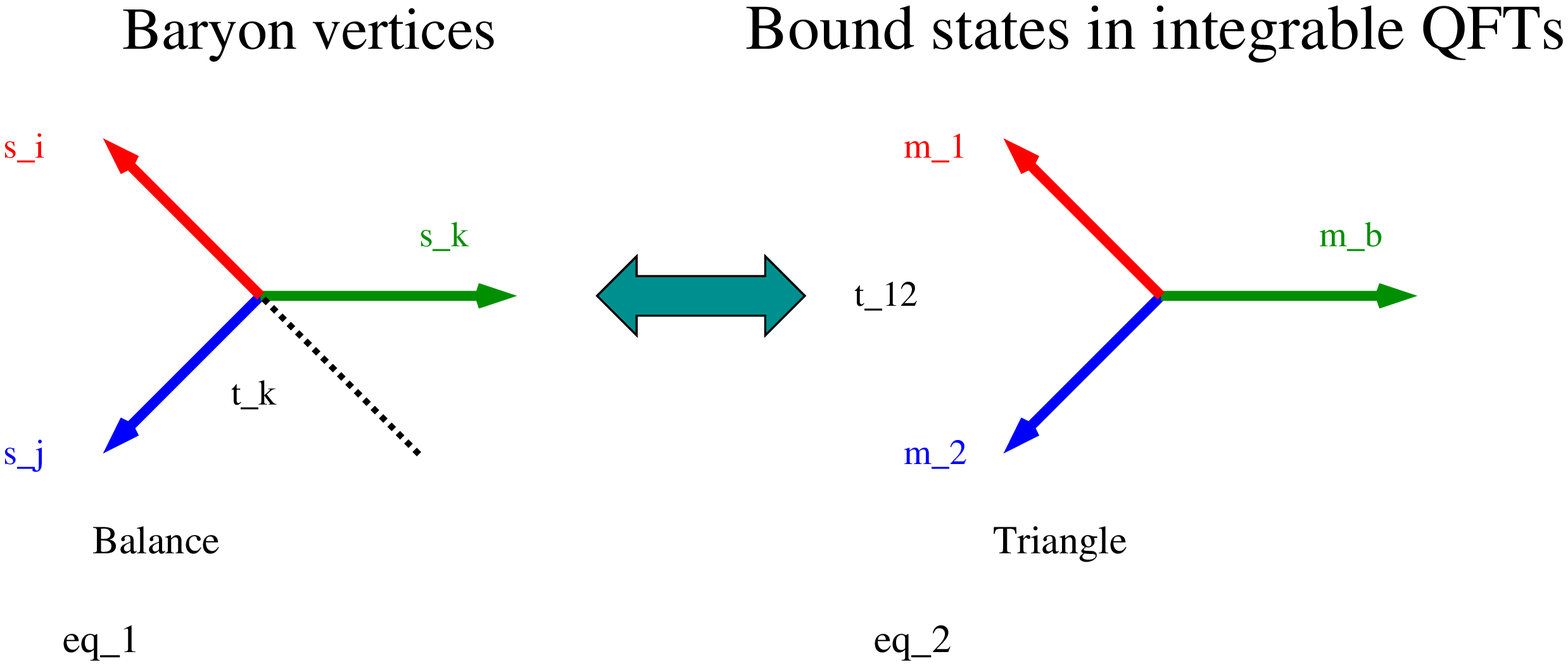} \caption{Pictorial representation of the relation between baryon vertices and mass spectrum.} 
\label{fig:baryon_mass} }

For such a process we know that $u_{\mathrm{S \bar S}}^{\\
\mathrm{B}_1}=\pi-\xi$ which, once inserted in (\ref{triangle}),  gives
\bea
M^2_1 = 2 M^2 (1- \cos \xi )  \ \ \ \  \to  \ \ \ \   \frac{M_1}{M} = 
2 \sin \frac{\xi}{2} \; ,
\eea
which is nothing but the mass formula used in the previous Section.

The crucial point is now to point out  that 
the balance of string tensions near $T=0$ and the mass triangle near 
$T_c$ have another common feature. In the former  the scaling properties 
of the baryon potential require that all the angles $\theta_k$ 
should not depend on $T$, therefore the string tension ratios 
$\sigma_k/\sigma$ stay constant as $T$ varies and define a line of constant physics starting at $T=0$. In the latter  the angles involved depend only on the marginal coupling $\xi$ hence a variation of the relevant coupling $\tau$   
of the 2D model (\ref{thermal}) generates a line of constant physics 
starting at $\tau=0$.

In the  gauge/CFT map established 
by the SY conjecture, in order to avoid a mismatch between the RG trajectories 
of the 3D gauge system and the corresponding 2D model, relevant 
perturbations of the CFT should correspond to relevant couplings of the 
critical gauge system. On the gauge side we pass from the $T\sim 0$ 
region to the $T\sim T_c$ by keeping constant the gauge couplings (hence
also the lattice spacing) and varying simply the size of the imaginary time direction. Thus the only physical parameter which is varied in passing from 
low temperature to $T_c$ is the {\sl reduced temperature} 
$t\equiv\frac{T-T_c}{T_c}$ of the gauge system, hence near $\tau\sim 0$
we have $\tau=\tau(t)$, while $\xi$ is kept constant.

Summing up, the variation of temperature of the gauge system defines a 
line of constant physics near $T=0$ and a similar line near $T=T_c$. 
The numerical work in the next two Sections will show that these two lines
are in fact a single one which goes through the whole confining phase.

\section{Monte Carlo setting and procedure}

\subsection{Mass ratio by correlators}

As introduced in Subsection~\ref{sub-operator}, 
we can determine the ratio $M_1/M$
using the large distance asymptotic behaviour of correlators; actually, 
exploiting the Svetitsky-Yaffe conjecture, we measured the 
Polyakov-Polyakov correlators $G_{\R}(R)$ of the (2+1)D $\mathbb{Z}_4$ gauge theory:
\eq
	G_{\R}(R)=\bra P_\R(0) P_\R^\dagger(R) \ket \; .
\en
In Section~\ref{dualref}, we have explained we can study this theory
by means of simulations on the dual 3D AT model and in~\cite{Giudice:2006hw}
the measurement of Polyakov-Polyakov correlators in both the fundamental
and double fundamental representations, $G(R)_{f}$ and $G(R)_{ff}$,
is described in detail.

Note that here, since we are studying the theory near the critical line,
the periodic boundary conditions play an important role; in this case there 
are 16 topologically different surfaces bounded by the two Polyakov lines
(4 for each Ising variable, as discussed in~\cite {Gliozzi:2005ny}),
therefore the correlator is the sum of these contributions. In fact, there
are only two important contributions, so we take into account only these two in our 
simulations (see Figures~\ref{fig:corr_f} and \ref{fig:corr_ff}).

\FIGURE{
\centering
\includegraphics[angle=0, width=10cm]{plot-corr-f.eps}
\caption{Polyakov-Polyakov correlator in the fundamental representation.}
\label{fig:corr_f}
}

\FIGURE{
\centering
\includegraphics[angle=0, width=10cm]{plot-corr-ff.eps}
\caption{Polyakov-Polyakov correlator in the double fundamental representation.}
\label{fig:corr_ff}
}

We have taken $10^6$ measures on the $64^2 \times 7$ lattice in the phase
space point $P$; $N_\tau = 7$ is chosen because it is the lowest possible
value above the deconfinement transition. Points on the plots are obtained
by independent simulations, one for each value of $R$ in the range 
$[15 \div 44]$.
These data are fitted using an expansion of the $K_0(mR)$ Bessel function,
truncated to first two terms,
\eq
G(R)=\mathrm{const} \times \frac{e^{-mR}}{\sqrt{mR}} \left[ 1- \frac{1}{8 m R} \right]
+ \mbox{``echo terms''},
\en 
in a range $[R_{min},R_{max}]$, where $R_{max}=44$; we have verified the 
results are stable when $R_{min}$ varies in the range $[22 \div 33]$.
Therefore, it is possible to determine the two masses:
\bea
a\,M_{ff} & = & 0.0698(15) \quad  (\chi^2/\mbox{d.o.f.} \approx 1.3) 
\nonumber \; , \\
a\,M_f    & = & 0.0433(8)  \quad  (\chi^2/\mbox{d.o.f.} \approx  1.2) \; ,
\eea
from which we can determine the ratio:
\eq
\sigma_2(T\sim T_c)/\sigma(T\sim T_c)=M_{ff}/M_f =1.612(46)\,.
\label{mffovermf}
\en
This result, obtained near the critical temperature, is compatible 
with the zero-temperature value (\ref{ratiosigma}), providing
a strong evidence for our conjecture.

\subsection{Estimating $\sigma_2/\sigma$ through the thermal  exponent $\nu$ with finite-size scaling}

To use the formula for the mass ratio, Eq.~(\ref{eq:ratio_nu}), we need a quite precise estimate for the 
thermal critical exponent $\nu$ in the phase space point $P$ (see Fig.~\ref{fig:atphasespace}). It can be 
obtained by means 
of a finite-size scaling analysis once the critical temperature for that 
choice of couplings has been identified.

The problem is, the system at the coupling $P$ turns out to be critical 
for a temperature $T_c$ such that $6 < \frac{1}{T_c} < 7$, hence, 
having to work with integer inverse temperatures, it is not possible 
to avoid some approximate method. The idea is then the following: 
choose a direction in the $(\alpha,\beta)$ phase space which crosses 
the critical line, and by moving from $P$ (in opposite directions) 
along it find two new points, $P_7$ and $P_6$, at which the system 
is  critical for temperatures $T=1/7$ and $T=1/6$ respectively. 
There, perform an estimate for $\nu$ with a standard finite-size scaling 
approach. Then, with a linear interpolation, construct the corresponding 
quantity for the original $P$ (see Fig.~\ref{fig:punti_mappa}).

\FIGURE{ \centering \includegraphics[angle=0, width=15cm]{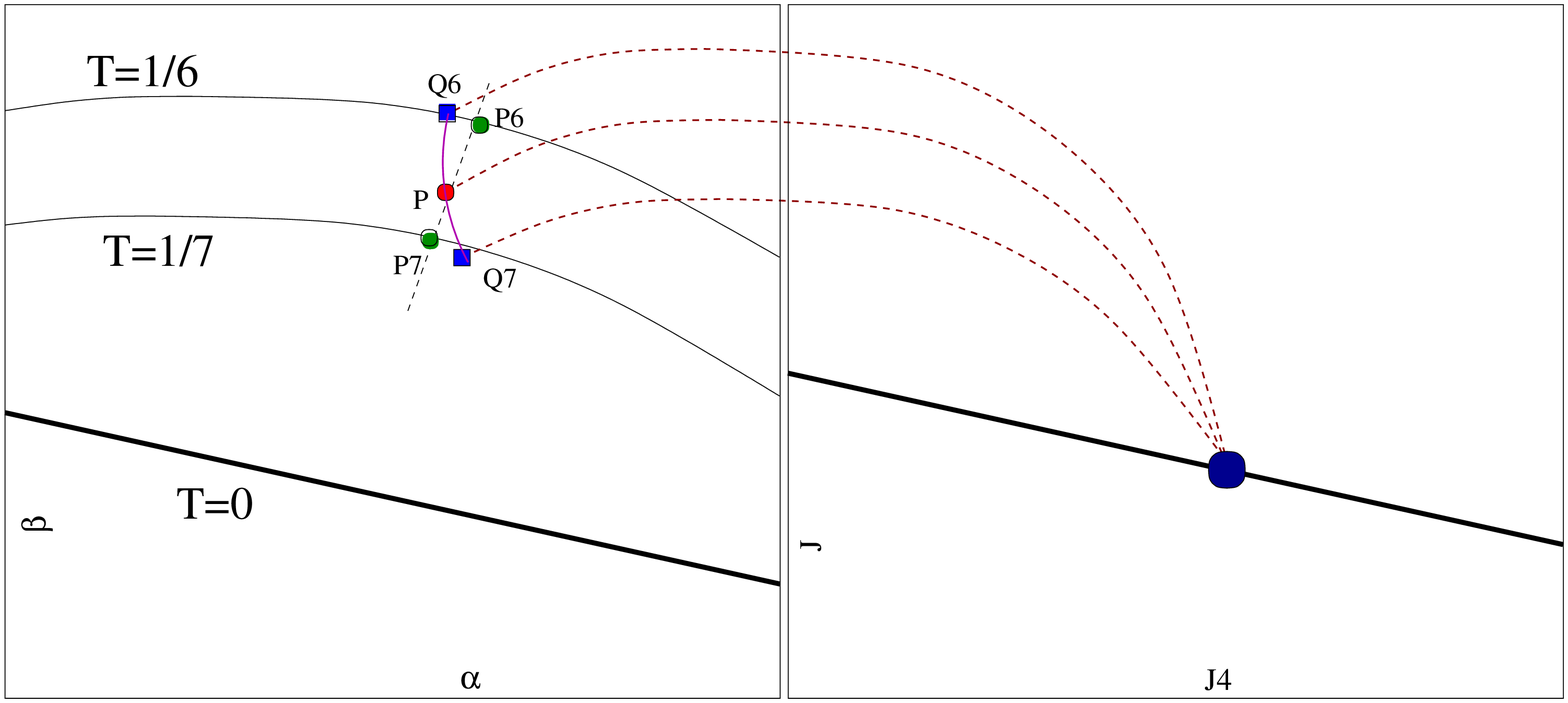} \caption{In 
the three-dimensional AT system (on the left), the point $P$ belongs to a whole 
line of points which exhibit the same critical behaviour, which is a RG 
trajectory in phase space (represented as a solid line passing through $P$). The 
intersections of this line with the finite-temperature critical lines for 
$\frac{1}{T}=6,7$ are labelled $Q6$ and $Q7$, but their exact position is 
unknown. Relying on a linear approximation, however, we located the points $P6, 
P7$ as described in the text, in order to identify, by interpolation, the point 
on the dimensionally-reduced AT model (on the right) to which $P$ is mapped.} 
\label{fig:punti_mappa} }

This first-order approximation, however, is motivated only if the
 points are close enough that the variation of $\nu$ is quite small; 
since the exact shape, in the phase space, of the trajectories of 
the RG (that is, the set of points which are mapped to the same point 
on the Gaussian model) is unknown, the best guess is to move from 
$P$ along a direction which is perpendicular to the zero-temperature 
critical line connecting the two  decoupled  Ising  and the 4-state 
Potts systems.

It has been shown long ago that in the SY context the plaquette operator 
is mapped into a combination of the unity and the energy operator of the 
corresponding CFT~\cite{Gliozzi:1997yc}.
Therefore, once the system is made  critical, one could extract the thermal 
exponent $\nu$ from the finite-size scaling behaviour of the plaquette
operator or some related observable that we denote with $\avg{\square}_L $,
where $L$ is the spacial size of the lattice.

In order to exploit the computational advantages of the dual 
transcription of the gauge model, it is convenient to evaluate directly 
the internal energy of the 3D AT model defined in (\ref{eq:3dataction}),
namely
\eq
\avg{\square}_L \equiv -\frac{1}{3L^2L_t} \avg{S_{AT}}\;\;.
\en
We expect the following finite-size critical behaviour as a 
function of the spacial side $L$ of the system:
\EQ
	\avg{\square}_L = \avg{\square}_\infty + b \cdot L^{\frac{1}{\nu}-d} \; ;
\EN
If one, instead, looks at the corresponding (density of) susceptivity:
\EQ
	\avg{\chi}_L \equiv \avg{ ( \square - \avg{\square}_L )^2 }_L \; ,
\EN
the power-law to compare with has the form:
\EQ
	\label{eq:suscettiva}
	\avg{\chi}_L = b' \cdot L^{\frac{2}{\nu}-d} \; ,
\EN
with the advantage that no constant additive terms are present, 
which could largely spoil the stability of the numerical results.

To locate the critical points $P_6$ and $P_7$, we proceeded in the phase space in a dichotomic way along the
 above-mentioned line from $P$ and looked for peaks in the plaquette susceptivity on a fixed spatial size 
 system. There is an intrinsic uncertainty on the exact critical values for $\alpha, \beta$, but in principle 
 it can be indefinitely shrunk by considering larger and larger lattices.

To perform the simulations, we used a cluster-based nonlocal update algorithm, an adaptation of the 
Swendsen-Wang prescription, based on alternating global updates on the two Ising subsystems in which the other
 variables play the role of a frozen background field. The algorithm is described in more detail in 
 \cite{confanomaly}.

We used $L=200$ finite-temperature lattices to find the couplings corresponding to $P_6$ and $P_7$, and for 
each sampled value of the couplings we took at least $\mathcal{O}(10^4)$ measurements. By locating the peak 
in the plaquette susceptibility (see Fig.~\ref{fig:peaks_susc}) we could identify the two points with a certain
 degree of accuracy as $P_6(0.0500965 \pm 0.0000063, 0.207235 \pm 0.000015)$ and $P_7(0.0497859 \pm 0.0000077,
  0.206478 \pm 0.000019)$.

\FIGURE{
	\centering
	\includegraphics[angle=0, width=10cm]{peaks_at_t7.eps}
	\caption{Behaviour of the plaquette susceptibility in the AT model at $T=1/7$ and spatial size $L=200$. 
	The clear peak allowed a precise estimate of the critical point.}
	\label{fig:peaks_susc}
}

Then, on exactly critical systems at $P_6$ and $P_7$, we took $\mathcal{O}(10^5)$ measurements of the plaquette 
at 26 values of spatial side $L$, ranging from $L=10$ to $L=165$. Not surprisingly, from the bare plaquette data 
the signal of the power-law behaviour was very noisy due to the presence of a constant background as another fit 
parameter, so we switched to using the prediction (\ref{eq:suscettiva}) for the susceptibility.

The data fitted very well to the expectation from $L=70$ already, so we could extract two values of the critical 
index $\nu$ (see Fig.~\ref{fig:susc_fss}):
\bea
	\nu_{T=1/6} & = & 0.8004(19)[22]\nonumber \; ; \\
	\nu_{T=1/7} & = & 0.7942(18)[38] \; ,
\eea
in which the first uncertainty refers to the statistical fluctuations while the second is an estimate of the 
systematic error in the measurement.

\FIGURE{
	\centering
	\includegraphics[angle=0, width=10cm]{susc_at_t7.eps}
	\caption{Finite-size scaling of the susceptibility of Eq.~\ref{eq:suscettiva} in the AT model at $T=1/7$ as a function of the system spacial size. The data allowed a precise estimate of the critical index $\nu$.}
	\label{fig:susc_fss}
}

By linear interpolation along the couplings, the value of $\nu$ and the (coupling-dependent) critical 
temperature $T_c$ was calculated for the very point $P$. It was found that $T_c(P) \simeq 0.1502 \simeq 
1/(6.655)$, which (since the value of $\sigma$ is well known for $P$) gives the universal ratio
\EQ
	\frac{T_c}{\sqrt{\sigma}} = 1.0393(12) \; .
\EN

 From the interpolation, we have $\nu(P) = 0.7984(19)[27]$. By plugging it into the formula for the mass ratio 
(\ref{eq:ratio_nu}), we obtain the following result:
\EQ
	\frac{M_1}{M}(P) = 1.6124(71)[102] \; ,
\label{mfromnu}
\EN
which is compatible with the less accurate estimate coming from the quantities in \cite{confanomaly} and 
thus well supports our conjecture.

\section{Conclusions}
In this paper we studied the ratio of the string tensions
$\sigma_2(T)/\sigma(T)$ near the deconfining point $T_c$ of a 3D
$\Z_4$ gauge model and compared the result with a general formula
which is expected to be true near $T=0$ for a generic gauge theory in
three or four dimensions. 

In this particular case we have combined numerical experiments with known 
exact results of  an integrable 2D quantum field theory that belongs, according to the Svetitsky-Yaffe 
conjecture, to the same universality class of the critical gauge system. 

An interesting property of the integrable model is that the mass ratio of the 
two physical states of the theory, which should equate the string tensions 
ratio near $T_c$, can be expressed as a simple function of the thermal 
exponent $\nu$  (see Eq.~(\ref{eq:ratio_nu})). Therefore we used two 
different methods to evaluate such a ratio: either a direct evaluation
of the string tensions through Polyakov loop correlators near $T_c$ 
(see Eq.~(\ref{mffovermf})) or  through a measure of $\nu$ 
(see Eq.~(\ref{mfromnu})). Both the estimates give compatible results which nicely agree with the ratio 
$\sigma_2/\sigma$ evaluated at $T=0$
(see Eq.~(\ref{ratiosigma})); for a schematic summary, see Table~\ref{tab1}.
We then conclude that, at least in this model, 
the k-string tensions ratios do not depend on $T$.

\begin{table}[htbp]
\begin{center}
\begin{tabular}{|l|ll|}
\hline
$\sigma_2(T)/\sigma(T)$ & Temperature & Method  \\
\hline
1.610(13)       & $T\simeq0+O(T^3)$  &  Eq.~(\ref{ratiosigma})  \\
1.612(46)       & $T \lesssim T_c$ &  Eq.~(\ref{mffovermf})    \\
1.6124(71)[102] & $T \lesssim T_c$ &  Eq.~(\ref{mfromnu})     \\
\hline
\end{tabular}
\end{center}
\caption{Numerical results.}
\label{tab1}
\end{table}

Even if in Section \ref{baryon} we gave  a general RG argument to support 
this assumption in a wider context,
we do not dare to extend such a conjecture to a general gauge system,
one reason being that if the deconfinement transition is first order,
as is the case in most gauge theories, we do not know a sound argument
to support it.


\begin{thebibliography}{99}

\bibitem{ds} M.~R.~Douglas and S.~H.~Shenker, \npb{447}{1995}{271}
[\hepth{9503163}].
\bibitem{hsz}  A.~Hanany, M.~J.~Strassler and A. Zaffaroni, \npb
{513}{1998}{87} [\hepth{9707244}].
\bibitem{hk} C.~P.~Herzog and I.~R.~Klebanov, \plb{526}{2002}{388}
[\hepth{0111078}].
\bibitem{as} A.~Armoni and M.~Shifman, \npb{671}{2003}{67}
[\hepth{0307020}].
\bibitem{kdk} F.~Gliozzi, \jhep{08}{2005}{063} [\hepth{0507016}].
\bibitem{bary} F.~Gliozzi, \prd{72}{2005}{055011} [\hepth{0504105}].
\bibitem{imam} Y.~Imamura, \ptp{115}{2006}{815} [\hepth{0512314}].
\bibitem{Armoni:2006ri}
  A.~Armoni and B.~Lucini,
  %``Universality of k-string tensions from holography and the lattice,''       
  \jhep{0606}{2006}{036}
  [\hepth{0604055}].
\bibitem{lt1}
B.~Lucini and M.~Teper, \plb{501}{2001}{128}
[\heplat{0012025}].
%%CITATION = HEP-LAT 0012025;%%                                                 

\bibitem{lt2}
B.~Lucini and M.~Teper, \jhep{06}{2001}{050}
[\heplat{0103027}].
%%CITATION = HEP-LAT 0103027;%%    


\bibitem{lt3}
B.~Lucini and M.~Teper, \prd{64}{2001}{105019}
[\heplat{0107007}].
%%CITATION = HEP-LAT 0107007;%%                                                 
\bibitem{lt4} B.~Lucini, M.~Teper and U.~Wegler, \jhep{04}{2004}{012}
[\heplat{0404008}].

\bibitem{kits}
Y.~Koma, E.~M.~Ilgenfritz, H.~Toki and T.~Suzuki,
%``Casimir scaling in a dual superconducting scenario of confinement,''         
\prd{64}{2001}{011501}
[\hepph{0103162}].
%%CITATION = HEP-PH 0103162;%%                                                  
\bibitem{ddprv1}
L.~Del Debbio, H.~Panagopoulos, P.~Rossi and E.~Vicari,
\prd{65}{2002}{021501}
[\hepth{0106185}].
\bibitem{ddprv2}
L.~Del Debbio, H.~Panagopoulos, P.~Rossi and E.~Vicari,
\jhep{01}{2002}{009}
[\hepth{0111090}].
\bibitem{ddpv}
L.~Del Debbio, H.~Panagopoulos and E.~Vicari,
\jhep{0309}{2003}{034}
[\heplat{0308012}].


\bibitem{confanomaly} P.~Giudice, F.~Gliozzi, S.~Lottini, 
JHEP \textbf{05} (2007) 010 [hep-th/0703153].

\bibitem{sy} B.~Svetitsky and L.~G.~Yaffe, Nucl.~Phys.~B {\bf 210} (1982) 423.

\bibitem{su4deforcrand} Ph.~de Forcrand, O.~Jahn, 
Nucl.~Phys.~{\bf 129} (Proc.~Suppl.) (2004) 709 [\heplat{0309153}].

\bibitem{dbgk} R.~V.~Ditzian, J.~R.~Banavar, G.~S.~Grest and L.~P.~Kadanoff,
\pr{22}{1980}{2542}.
\bibitem{az} P.~Arnold and Y.~Zhang, \npb{501}{1997}{803} [\heplat{9610032}].



\bibitem{baxter}
R.~J.~Baxter, \emph{Exactly solved models in statistical mechanics} 
(Academic Press, New York, 1982). 

\bibitem{kadanoff}
L.~P.~Kadanoff and A.~C.~Brown, 
Ann.~Phys.~121 (1979) 318.

\bibitem{revgius}
G.~Mussardo, Phys.~Rept.~{\bf 218} (1992) 215; 
A.~B.~Zamolodchikov and Al.~B.~Zamolodchikov, 
Ann.~Phys.~120 (1979) 253. 

\bibitem{aldopaolo}
G.~Delfino, Phys.~Lett.~B 450 (1999) 196; 
G.~Delfino and P.~Grinza, Nucl.~Phys.~B {\bf 682} (2004) 521 
[\hepth{0309129}].


\bibitem{refsuSGsmatrix}
A nice introduction to the Sine-Gordon model and its S-matrix 
can be found here:  
R.~Rajaraman,
\emph{Solitons And Instantons. An Introduction To Solitons And Instantons In
Quantum Field Theory},
%\href{http://www.slac.stanford.edu/spires/find/hep/www?irn=998575}{SPIRES entry}
Amsterdam, Netherlands: North-holland (1982) 409p.

\bibitem{smiyuzam}
V.~P.~Yurov and A.~B.~Zamolodchikov, 
Int.~J.~Mod.~Phys.~A {\bf 6} (1991) 3419; 
F.~A.~Smirnov, \emph{Form Factors in Completely Integrable 
Models of Quantum Field Theory} (World Scientific) 1992. 

\bibitem{chp02}
  M.~Caselle, M.~Panero and P.~Provero,
  %``String effects in Polyakov loop correlators,''
  JHEP {\bf 0206} (2002) 061
  [\heplat{0205008}].
  %%CITATION = JHEPA,0206,061;%%

\bibitem{bc05}
  M.~Bill\`o and M.~Caselle,
  %``Polyakov loop correlators from D0-brane interactions in bosonic string
  %theory,''
  JHEP {\bf 0507} (2005) 038
  [\hepth{0505201}].
  %%CITATION = JHEPA,0507,038;%%

\bibitem{z3potts}
M.~Caselle, G.~Delfino, P.~Grinza, O.~Jahn and N.~Magnoli,
%``Potts correlators and the static three-quark potential,''                    
J.~Stat.~Mech.~{\bf 0603} (2006) P008
[\hepth{0511168}].

\bibitem{eden}
R.~J.~Eden, P.~V.~Landshoff, D.~I.~Olive, and J.~C.~Polkinghorne,
{\it The Analytic S-Matrix}, Cambridge Univ. Press, 1966.

%\cite{Giudice:2006hw}
\bibitem{Giudice:2006hw}
P.~Giudice, F.~Gliozzi and S.~Lottini,
%``Quantum broadening of k-strings in gauge theories,''
JHEP {\bf 0701} (2007) 084
[\hepth{0612131}].
%%CITATION = JHEPA,0701,084;%%

%\cite{Gliozzi:2005ny}
\bibitem{Gliozzi:2005ny}
F.~Gliozzi, S.~Lottini, M.~Panero and A.~Rago,
%``Random percolation as a gauge theory,''
Nucl.~Phys.~B {\bf 719} (2005) 255
[\condmat{0502339}].
%%CITATION = NUPHA,B719,255;%%

\bibitem{Gliozzi:1997yc}
  F.~Gliozzi and P.~Provero,
  %``The Svetitsky-Yaffe conjecture for the plaquette operator,''
  Phys.~Rev.~D {\bf 56} (1997) 1131
  [\heplat{9701014}].
  %%CITATION = PHRVA,D56,1131;%%


\end{thebibliography}
\end{document}